\documentclass[a4paper,11pt]{article}
\usepackage{jcappub}
\usepackage[T1]{fontenc}
\usepackage{%
amsmath,
amsthm,
amssymb,
braket,
bm,
aas_macros
}
\usepackage{color}
\title{Limits on primordial magnetic fields from primordial black hole abundance}
\author[a]{Shohei Saga,}
\author[b]{Hiroyuki Tashiro,}
\author[c,d]{and Shuichiro Yokoyama}

\affiliation[a]{Center for Gravitational Physics, Yukawa Institute for Theoretical Physics, Kyoto University, Kyoto 606-8502, Japan}
\affiliation[b]{Department of Physics and Astrophysics, Nagoya University, Nagoya, 464-8602, Japan}
\affiliation[c]{Kobayashi Maskawa Institute, Nagoya University, Aichi 464-8602, Japan}
\affiliation[d]{Kavli Institute for the Physics and Mathematics of the Universe (WPI),
Todai institute for Advanced Study, University of Tokyo, Kashiwa, Chiba 277-8568, Japan}

\emailAdd{shohei.saga@yukawa.kyoto-u.ac.jp}
\arxivnumber{2002.01286}
\abstract{
Primordial magnetic field (PMF) is one of the feasible candidates to explain observed large-scale magnetic fields, for example, intergalactic magnetic fields.
We present a new mechanism that brings us information about PMFs on small scales based on the abundance of primordial black holes (PBHs).
The anisotropic stress of the PMFs can act as a source of the super-horizon curvature perturbation in the early universe.
If the amplitude of PMFs is sufficiently large, the resultant density perturbation also has a large amplitude, and thereby, the PBH abundance is enhanced.
Since the anisotropic stress of the PMFs is consist of the square of the magnetic fields, the statistics of the density perturbation follows the non-Gaussian distribution.
Assuming Gaussian distributions and delta-function type power spectrum for PMFs, based on a Monte-Carlo method, we obtain an approximate probability density function of the density perturbation, and it is an important piece to relate the amplitude of PMFs with the abundance of PBHs.
Finally, we place the strongest constraint on the amplitude of PMFs as a few hundred nano-Gauss on $10^{2}\;{\rm Mpc}^{-1} \leq k\leq 10^{18}\;{\rm Mpc}^{-1}$ where the typical cosmological observations never reach.
}
\begin{document}
\begin{flushright}
YITP-20-10\\
\end{flushright}
\maketitle
\flushbottom

\section{Introduction}
Magnetic fields are ubiquitous in the Universe.
Recent observations of high-energy TeV photons emitted by distant blazars imply the existence of large-scale magnetic fields, especially, the intergalactic and/or void magnetic fields (see e.g., Refs.~\cite{2010Sci...328...73N,2011MNRAS.414.3566T,2012ApJ...747L..14V,2013ApJ...771L..42T,2015RAA....15.2173Y,2017ApJ...847...39V} and references therein).
Through these observations, for example, Ref.~\cite{2010Sci...328...73N} reported the lower bound on intergalactic magnetic fields as $3\times 10^{-16}\; {\rm Gauss}$ with their coherent length, $\lambda_{B} \gtrsim 0.1\; {\rm Mpc}$ (at scales below $0.1\; {\rm Mpc}$, the lower bound gets stronger as proportional to $\lambda_{B}^{-1/2}$).
While the lower bounds based on the above observations depend on the coherent length of magnetic fields, it has been recognized that the evidences of large-scale magnetic fields are supported.
Although the origin of the intergalactic magnetic fields has not been revealed yet, a bunch of theories on the generation of large-scale magnetic fields are proposed (see e.g., Refs.~\cite{2013A&ARv..21...62D,2016RPPh...79g6901S,2001PhR...348..163G} for reviews).

The generation mechanism of magnetic fields are classified into two categories.
One is the astrophysical scenario, which is driven by the dynamics of the astrophysical objects~(e.g., Refs.~\cite{2000ApJ...540..755D,2005ApJ...633..941H,2013PhRvL.111e1303N,2005A&A...443..367L}).
This scenario seems to work well for small-scale magnetic fields, below galactic scales.
However it would be difficult to explain large-scale magnetic fields such as magnetic fields in galaxy clusters and large-scale filaments, because of the limited both length and time scales for the generation mechanism.
Another scenario for large-scale magnetic fields is the primordial origin in which the seed field, namely, primordial magnetic fields (PMFs), is generated in the early universe, especially before the epoch of cosmic recombination.
Since the generation mechanism owes to cosmological phenomena, the resultant PMFs could be on cosmological scales.
Therefore, the PMF scenario is attractive for the origin of observed large-scale magnetic fields.
In addition, interestingly, observed intergalactic (void) magnetic fields might be considered as the direct evidence of PMFs.
The future analysis of the Faraday Rotation Measure from Fast Radio Bursts would distinguish whether observed magnetic fields are the initial seed origin or astrophysical origin~\cite{2017CQGra..34w4001V,2018MNRAS.480.3907V}.

One of the most interesting scenarios for the generation of the primordial magnetic fields is the inflationary magnetogenesis.
In such a scenario, the super-horizon scale magnetic fields can be generated from quantum fluctuations as well as curvature perturbations or primordial gravitational waves.
Since Maxwell's theory is invariant under the conformal transformation, the success of inflationary magnetogenesis is required new interaction between electromagnetic fields and any other fields which breaks a conformal invariance, e.g., Refs.~\cite{1988PhRvD..37.2743T,1992ApJ...391L...1R,1993PhRvD..48.2499D,2004PhRvD..69d3507B,2009JCAP...08..025D,2011PhR...505....1K,2014JCAP...10..056C}.
Other possibilities of primordial magnetogenesis
have also discussed even in the post-inflation era, e.g., based on the Harrison mechanism~\cite{1970MNRAS.147..279H,2005PhRvL..95l1301T,2011MNRAS.414.2354F,2015PhRvD..91l3510S,2016PhRvD..93j3536F}, or the bubble collision/turbulence in the cosmological phase transitions~\cite{1991PhLB..265..258V,1997PhRvD..55.4582S,2012ApJ...759...54T,2019arXiv190704315E}.
While many authors have been tackled to construct a magnetogenesis model, there are a few candidates for the succeeded model, e.g. Ref.~\cite{2016EL....11519001D}.
To not only distinguish models but also to obtain a hint of generation
mechanism and physics in the early universe, we need to survey the effects of PMFs on the cosmological observations seriously.

The effects of PMFs on cosmological observations have been discussed in many works of literature.
For instance, if PMFs exist during the Big Bang Nucleosynthesis (BBN) epoch, the abundance of the light elements is heavily modified~\cite{2002PhRvD..65b3517C,2012PhRvD..86l3006Y,2012PhRvD..86f3003K,2019ApJ...872..172L}.
The upper limit from BBN on the present strength of PMFs is $1.5\times 10^{3}\; {\rm nG}$~\cite{2012PhRvD..86f3003K}.
PMFs can be a source of the CMB temperature and polarization anisotropies in various ways~\cite{2004PhRvD..70d3011L,2004PhRvD..70l3507G,2008PhRvD..78b3510F,2009MNRAS.396..523P,2010PhRvD..81d3517S,2013PhRvD..88h3515B,2016A&A...594A..19P,2019arXiv190409121S},
large-scale structure of the Universe~\cite{2012SSRv..166....1R,2012JCAP...11..055F,2012PhRvD..86d3510S,2014JCAP...03..027C}, and in particular, the detailed analysis of the magnetohydrodynamic effect on CMB temperature anisotropy gives an upper limit on the amplitude of PMFs as $47\times 10^{-3} \; {\rm nG}$ \cite{2019PhRvL.123b1301J}.
Note that the upper limit from CMB anisotropy is sensitive to PMFs at $O({\rm Mpc})$ sales.
PMFs release their energy via the dissipation into the CMB photons, and the injected energy distorts the CMB spectrum~\cite{1998PhRvD..57.3264J,1998PhRvD..58h3502S}.
The upper limit from the CMB spectral distortion is $30\; {\rm nG}$ on the dissipation scale of PMFs~\cite{2000PhRvL..85..700J,2014JCAP...01..009K}.
Alternatively, the upper limit on PMFs by using 21cm radiation has been discussed in Refs.~\cite{2009ApJ...692..236S,2009JCAP...11..021S,2019MNRAS.488.2001M,2019JCAP...01..033K}.
Since the anisotropic stress of PMFs induces gravitational waves, direct detection of gravitational wave background can constrain the amplitude of PMFs on a very small scale through the Pulsar Timing Arrays (PTAs)~\cite{2000PhRvD..61d3001D,2018PhRvD..98h3518S}.
Thus, even though the upper limit on large-scale PMFs is well established by CMB observations, that on small-scale PMFs is still developing.

In this paper, we propose a new method to constrain relatively small-scale PMFs based on primordial black hole (PBH) abundance.
The PBH is formed through the gravitational collapse of an overdense region in the early Universe, and such overdense region can stochastically exist due to the primordial curvature fluctuations with a large amplitude \cite{1967SvA....10..602Z,Hawking:71,Carr:75,CarrHawking:74}. 
In particular, in the radiation-dominated Universe, such a PBH formation would occur at the time when the size of the overdense region enters the Hubble horizon.
Then, the formation probability of PBHs, directly corresponding to the PBH abundance, depends on the statistical property of the primordial fluctuations on super-horizon scales. 
In particular, PBH might be formed at a rare peak, and then the formation probability is sensitive to the tail of the probability density function (PDF) of the primordial fluctuations.
Still, there is no strong evidence of the existence of PBHs, and we have only upper limits for their abundance obtained from various observations (see e.g., Refs.~\cite{Carr:2009jm, Sasaki:2018dmp} and references therein).
The observational constraints on the PBH abundance are translated into the upper bound on the amplitude of the primordial density fluctuations (see e.g., Ref. \cite{Sato-Polito:2019hws}), also with taking their non-Gaussian property into account (see e.g., Ref. \cite{Nakama:2017xvq}).

Incidentally, the anisotropic stress of PMFs can also induce the primordial curvature perturbation on super-horizon scales before the neutrino decoupling.
When the amplitude of PMFs is sufficiently large, the induced curvature perturbations also have large amplitude, and subsequently, PBHs are overproduced.
In principle, we can place the upper limit on the amplitude of PMFs through the resultant abundance of PBHs.
To do so, we have to carefully study the PDF of the primordial curvature perturbations induced by PMFs.
Even if PMFs obey the exact Gaussian statistics, the anisotropic stress of the PMFs is given by the square of the magnetic field, and hence, the sourced curvature/density fluctuations must follow non-Gaussian statistics.
To obtain the PDF of the sourced density fluctuations, we perform the Monte-Carlo method by following Ref.~\cite{2016PhRvD..94d3507N}.
The relation between the amplitude of PMFs and the PBH abundance is firstly investigated in the following sections.

This paper is organized as follows.
In Sec.~\ref{sec: passive mode}, we discuss the statistical properties of the density perturbation induced by PMFs.
In particular, since such density perturbations obey highly non-Gaussian distribution, we use the Monte-Carlo method to obtain the PDF based on Ref.~\cite{2016PhRvD..94d3507N}.
In Sec.~\ref{sec: result}, we relate the PBH abundance to the amplitude of PMFs by assuming the delta-function type power spectrum.
With the current upper limit on the PBH abundance, we give the upper limit on PMFs.
Finally, we devote Sec.~\ref{sec: summary} to the summary of this paper.
Throughout the paper, we will work in units $c=\hbar=1$.

\section{Super-horizon primordial curvature perturbations from PMFs}
\label{sec: passive mode}

In this section, we discuss the statistical property of the initial density perturbations induced by PMFs, which strongly affects the PBH abundance.
In particular, the PBH abundance is related to the probability density function~(PDF) of the density fluctuations.
As we will see later, such density perturbation does not obey the Gaussian distribution, and as a result, the resultant abundance of PBHs is non-trivial.

It is known that PMFs create two kinds of curvature perturbation; passive magnetic mode and compensated magnetic mode.
The passive mode arises on both super- and sub-horizon scales, while the compensated mode is generated on only sub-horizon scales~\cite{2010PhRvD..81d3517S,2012PhRvD..86d3510S}.
In the PBH formation during the radiation-dominated era, the perturbation on the horizon-crossing scale is crucial.
Therefore, it is sufficient to focus on only the passive mode.
In the following section, first of all, we review the curvature perturbation generated from PMFs during the radiation dominated era~\cite{2010PhRvD..81d3517S}.

\subsection{Passive curvature perturbation}

Before the neutrino decoupling, the non-negligible anisotropic stress of PMFs contributes to the total energy-momentum tensor of the Universe.
Although it is compensated by arising the anisotropic stress of neutrinos after the neutrino decoupling, the scalar part of the anisotropic stress due to PMFs is a source of the curvature perturbation.
Resultantly, the curvature perturbation can grow between the epochs of the PMF generation and the neutrino decoupling even on super-horizon scales.
At a conformal time $\eta$ during this growing regime, the super-horizon ($k\eta \ll 1$) curvature perturbation on comoving slice induced by the anisotropic stress of PMFs is given in Fourier space as \cite{2010PhRvD..81d3517S}
\begin{eqnarray}
\zeta_{B}(\bm{k},\eta) = - \frac{1}{3}\xi(\eta) R_{\gamma}\Pi_{B}(\bm{k})
 ~, \label{eq:zetab}
\end{eqnarray}
with
\begin{eqnarray}
\xi(\eta) =
\begin{cases}
 \log \left( \frac{\eta}{\eta_B} \right) + \frac{\eta_B}{2 \eta} - \frac{1}{2} & (\eta_B < \eta < \eta_\nu) \\
 \log \left( \frac{\eta_\nu}{\eta_B} \right) + \left( \frac{5}{8 R_\nu} - 1 \right) & (\eta > \eta_\nu) 
\end{cases}
 ~, \label{eq:xiform}
\end{eqnarray}
where $R_{i}$ represents an energy fraction of the $i$-component in the total radiation defined as $R_i \equiv \rho_{i}/\sum \rho_i$ ($i = \gamma$ (photon), $\nu$ (neutrinos)) and
$\eta_{\nu}$ and $\eta_B$ respectively represent the neutrino decoupling and the PMF generation epochs in terms of the conformal time.
In the equation~\eqref{eq:zetab}, $\Pi_{B}$ represents the scalar part of the anisotropic stress of PMFs, which is defined as
\begin{equation}
\Pi_{B}(\bm{k}) =
\left( \hat{k}_{i}\hat{k}_{j} - \frac{1}{3}\delta_{ij}\right)
\frac{9}{8\pi \rho_{\gamma,0}} \int\frac{{\rm d}^{3}k_{1}}{(2\pi)^{3}}\; 
B_{i}(\bm{k}_{1}) B_{j}(\bm{k}-\bm{k}_{1})
~. \label{eq: def Pi_{B}}
\end{equation}
Here $\rho_{\gamma,0}$ is the photon energy density at the present, and $\bm{B}(\bm{k})$ is the Fourier component of comoving PMFs, which corresponds to the PMFs at the present epoch after evolving adiabatically since the generation as proportional to $\propto a^{-2}$ with a scale factor $a$.

Since we suppose that $\bm{B}(\bm{k})$ is a random Gaussian field, the statistical property of the PMF $\bm{B}(\bm{k})$ is completely
given by the power spectrum
\begin{equation}
\Braket{B_{i}(\bm{k})B_{j}(\bm{k'})} = (2\pi)^{3}\delta^{3}_{\rm D}(\bm{k}+\bm{k'})\frac{\delta_{ij}-\hat{k}_{i}\hat{k}_{j}}{2} P_{B}(k) ~,
 \label{eq:twopoint}
\end{equation}
where $\delta^{3}_{\rm D}(\bm{k})$ is a Dirac delta function and $P_{B}(k)$ stands for a power spectrum of PMFs.
In the followings, we investigate the statistical property of the density perturbations from the passive curvature perturbation based on Eq.~\eqref{eq:twopoint}.

\subsection{Density perturbations and their statistical properties}

The PBH formation is relevant to the density perturbation at the horizon crossing scale.
On comoving slice, up to the linear order the connection between the density perturbation and the curvature perturbation during the radiation dominated era is given by
\begin{equation}
\delta_{B}(\bm{k},\eta)
= \frac{4}{9}\left( \frac{k}{aH}\right)^{2} 
\zeta_{B}(\bm{k},\eta) ~,
\label{eq: delta and zeta}
\end{equation}
where $H$ and $a$ are a Hubble parameter and a scale factor at $\eta$, respectively.
During the radiation-dominated era, it can be roughly considered that PBHs are formed by the gravitational collapse of overdense region soon after the size of the overdense region reenters the horizon.
Thereby, the mass of the formed PBH is characterized by the horizon scale at that moment.

We introduce the smoothed density perturbations over the comoving radius $R$ given by
\begin{eqnarray}
\delta_{B}(\bm{x},\eta; R)
&=& \int{\rm d}^{3}x'\; W_{R}(|\bm{x}-\bm{x}'|)\delta_{B}(\bm{x}',\eta) \notag \\
&=& \int\frac{{\rm d}^{3}k}{(2\pi)^{3}}\; W(kR) \delta_{B}(\bm{k},\eta) e^{i\bm{k}\cdot\bm{x}} ~, \label{eq: delta R x}
\end{eqnarray}
where, we use the Gaussian window function for smoothing in this paper:
\begin{equation}
W(x) = e^{-x^{2}/2} ~.
\end{equation}
Later, we will take the smoothing scale $R$ to be the comoving horizon scale at the PBH formation in the evaluation of the PBH abundance.

\subsubsection{Mean and dispersion}

In this section, let us discuss the statistical property of the smoothed density perturbation given in Eq.~(\ref{eq: delta R x}).
First, we show the mean value of the density perturbation.
With Eqs.~\eqref{eq:zetab}--\eqref{eq: def Pi_{B}}, the mean value of the curvature perturbation induced by PMFs in the real space vanishes, $\Braket{\zeta_{B}(\bm{x})} = 0$, and thereby, the mean of density perturbation is zero.
This is due both to the divergence free nature of magnetic fields and to the property of the projection tensor on the scalar mode.
Note that this result is different from the property of the curvature perturbation induced by the second-order primordial gravitational waves discussed in Refs.~\cite{2015PhRvD..92l1304N,2016PhRvD..94d3507N}.

\begin{figure}[t]
\centering
\includegraphics[width=0.6\textwidth]{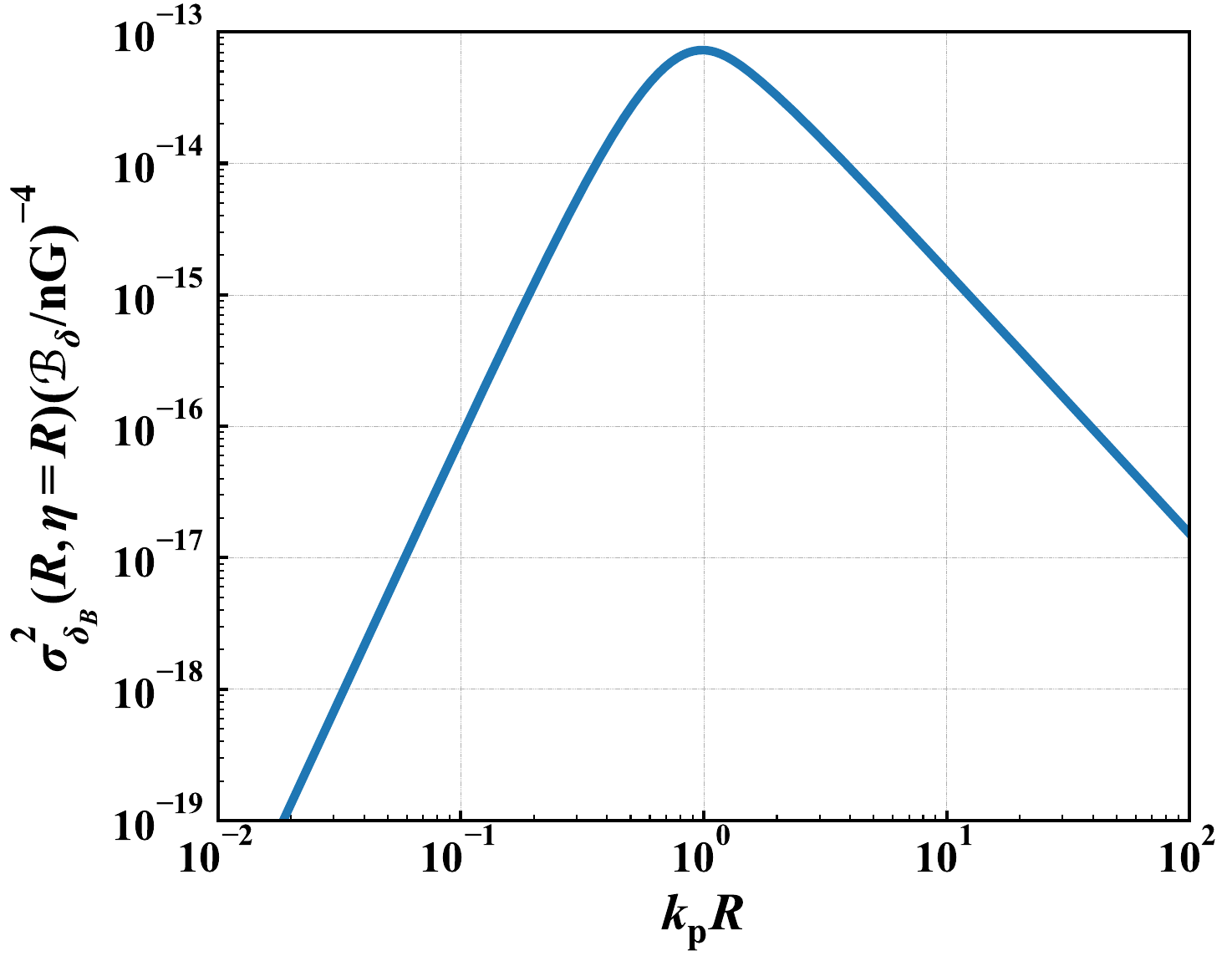}
\caption{
Dispersion normalized by the amplitude of PMFs $\sigma^{2}_{\delta_{B}}(R,\eta=R) \left(\mathcal{B}_{\delta}/{\rm nG} \right)^{-4}$ given in Eq.~(\ref{eq: dispersion delta-type}).
The peak wavenumber is numerically obtained as $k_{\rm p}R\approx 0.9872$.
}
\label{fig: dispersion}
\end{figure}

Next, we focus on the dispersion of the density perturbations.
According to Eqs.~(\ref{eq: delta and zeta}) and (\ref{eq: delta R x}), in terms of the primordial curvature perturbations, the dispersion is given by\footnote{
Here, we neglect the transfer function which describes the evolution of the curvature perturbations on sub-horizon scales.
For relatively broad power spectrum, the sub-horizon modes can contribute the smoothed density perturbations, and the transfer function should be included in the estimation of the dispersion. Furthermore, the choice of the window function might affect the PBH abundance~\cite{Ando:2018qdb,Young:2019osy}. However, here, as we will show later we focus on the delta-function type power spectrum and in such a case the contribution from the sub-horizon modes is safely neglected.
}
\begin{eqnarray}
\sigma^{2}_{\delta_{B}}(R,\eta) &=&
\Braket{ \left( \delta_{B}(\bm{x},\eta; R) \right)^{2}} \notag \\
&=&
\frac{16}{81}
\int\frac{{\rm d}k}{k}
W^{2}(kR)
\left( k\eta\right)^{4}
\mathcal{P}_{\zeta_{B}}(k,\eta) ~, \label{eq: dispersion}
\end{eqnarray}
where we define the dimensionless power spectrum,~$\mathcal{P}_{\zeta_B}(k,\eta)$, of the curvature perturbation as
\begin{equation}
\Braket{\zeta_{B}(\bm{k},\eta)\zeta_{B}(\bm{k'},\eta)} =
(2\pi)^{3}\delta^{3}_{\rm D}(\bm{k} + \bm{k'}) \frac{2\pi^{2}}{k^{3}}\mathcal{P}_{\zeta_{B}}(k,\eta) ~. \label{eq: PS zeta def}
\end{equation}
Combining Eqs.~(\ref{eq:zetab}), (\ref{eq: def Pi_{B}}), and (\ref{eq: PS zeta def}), the dimensionless power spectrum of the passive curvature perturbation is obtained as
\begin{equation}
\frac{2\pi^{2}}{k^{3}}\mathcal{P}_{\zeta_{B}}(k,\eta)
= \frac{1}{18}\xi^{2}(\eta) R^{2}_{\gamma}
\left( \frac{9}{8\pi \rho_{\gamma,0}} \right)^{2}
\frac{1}{(2\pi)^{2}}
\int{{\rm d}k_{1}}\; k^{2}_{1} P_{B}(k_{1})
\int^{1}_{-1}{{\rm d}\mu}\; 
\mathcal{I}(k,k_{1},\mu) P_{B}(|\bm{k}-\bm{k}_{1}|) ~,
\end{equation}
where $\mu = \hat{\bm{k}}\cdot \hat{\bm{k}}_{1}$ and we define the configuration kernel as
\begin{eqnarray}
\mathcal{I}(k,k_{1},\mu)
& = &
\frac{1}{9}\frac{k^{2}(1+\mu^{2}) + k k_{1}(2\mu - 6\mu^{3}) + k^{2}_{1}(5-12\mu^{2}+9\mu^{4})}{k^{2} + k^{2}_{1} - 2k k_{1} \mu} ~.
\end{eqnarray}
Throughout this paper, for simplicity, we focus on the delta-function type power spectrum for PMFs given by
\begin{equation}
P_{B}(k) = \frac{2\pi^{2}}{k^{3}}\mathcal{B}^{2}_{\delta} k_{\rm p} \delta_{\rm D}(k-k_{\rm p}) ~. \label{eq: PS def}
\end{equation}
Then, we obtain the dispersion of the smoothed density perturbation for the delta-function type power spectrum as
\begin{align}
\sigma^{2}_{\delta_{B}}(R,\eta)
&\approx
1.12\times 10^{-16} \xi^{2}(\eta) R^{2}_{\gamma}\left( \frac{\mathcal{B}_{\delta}}{1\; {\rm nG}}\right)^{4}
\frac{\left( k_{\rm p} \eta\right)^{4}}{(k_{\rm p}R)^{10}}
\notag \\
&\times
\Bigl[
(3-12(k_{\rm p}R)^{2} + 20 (k_{\rm p}R)^{4})
+e^{-4(k_{\rm p}R)^{2}}(-3 + 4(k_{\rm p}R)^{4} - 16(k_{\rm p}R)^{6} -64 (k_{\rm p}R)^{8})
\Bigr]
~.
\label{eq: dispersion delta-type}
\end{align}
One of the crucial physical quantities for the PBH formation is the dispersion of the density perturbations smoothed over the horizon scale, that is, $R \sim (aH)^{-1} = \eta$, evaluated at the horizon crossing time.
Taking $\eta=R$ in Eq.~\eqref{eq: dispersion delta-type}, we obtain the dispersion $\sigma^{2}_{\delta_{B}}(R,\eta=R)$ as a function of only $k_{\rm p}R$ and plot it in Fig.~\ref{fig: dispersion}.
From Fig.~\ref{fig: dispersion}, the dispersion has a peak at $k_{\rm p }R\approx 0.9872$.

This fact enables us to conclude that, when the PMF has the delta-function power spectrum with the peak wavenumber $k_{\rm p}$, PBHs are mainly produced by the corresponding conformal time $\eta = R \approx 0.9872/k_{\rm p}$, which is roughly equivalent to the horizon crossing time for the scale $k = k_{\rm p}$.
We will therefore focus on the PBH formation at only this conformal time satisfying $\eta_{\rm p} = 0.9872/k_{\rm p}$ for different $k_{\rm p}$.

\subsubsection{Probability density functions}\label{sec: PDF}
\begin{figure}[t]
\centering
\includegraphics[width=0.6\textwidth]{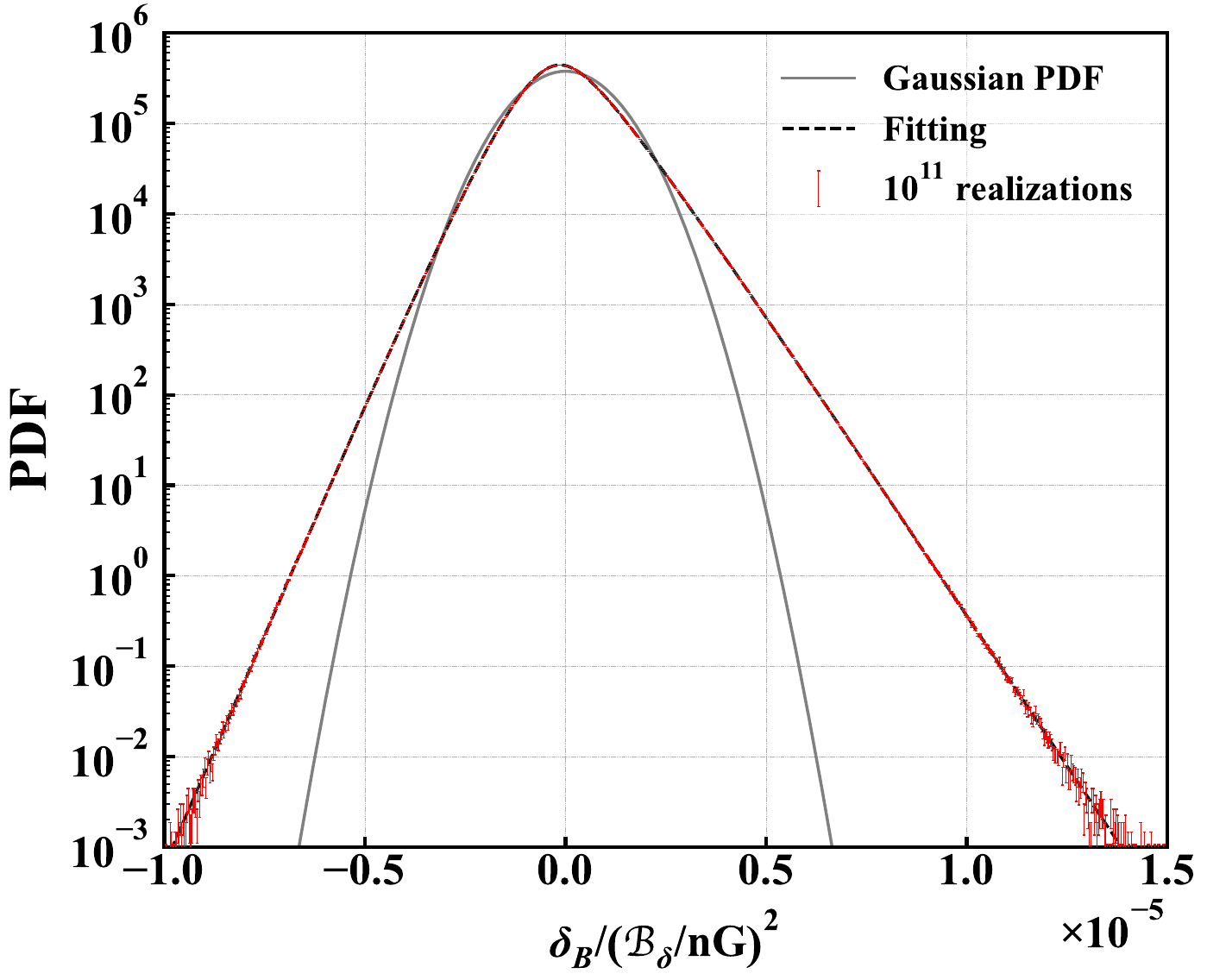}
\caption{
Probability density functions obtained by nearly $10^{11}$ realizations.
Just for information, we show the error bars which corresponds to the $1\sigma$ Poisson error.
The gray line is a Gaussian PDF with the same dispersion.
}
\label{fig: PDF}
\end{figure}

For the Gaussian density perturbations, the PBH abundance
should be determined only by the ratio between the critical density and the dispersion of the density perturbations.
However, since the statistics of the density perturbations induced by PMFs is highly non-Gaussian, we should discuss their PDF for the more precise estimation of the PBH abundance.
We use the Monte-Carlo method by following Ref.~\cite{2016PhRvD..94d3507N} to estimate the PDF of the smoothed density perturbation.
The details of the method are presented in Appendix.~\ref{sec: app PDF}.
Based on Appendix.~\ref{sec: app PDF}, we perform approximately $10^{11}$ realizations and obtain the PDF shown in Fig.~\ref{fig: PDF}.
Since we are not interested in the mathematically exact form of the PDF in order to obtain the upper limit on PMFs, we find and use the asymptotic behavior as the black dashed line in Fig.~\ref{fig: PDF}, which is given by a linear in log-space as $P_{\rm MC}(x= \delta_{B}/\mathcal{B}^{2}_{\delta}) \propto e^{a |x|}$ for $0\ll |x|$.
We use this asymptotic fitting function on $0\ll |x|$.
On the other hand, the intermediate part, $x\approx 0$, is interpolated by using the central value of the numerical realization.

For comparison, we plot the Gaussian distribution with the same dispersion $\sigma\approx 1.06\times 10^{-6}$ as in Eq.~\eqref{eq: dispersion delta-type} in the gray line.
The effect of the non-Gaussian effect appears on the tail part in both the overdensity and underdensity side.

\section{Results}\label{sec: result}

In this section, following the previous section we evaluate the PBH abundance, and based on the result, we show our main results of the upper limit on PMFs.

\subsection{Abundance of PBHs}

For the PDF of the density field $P(\delta)$, the fraction of the energy density collapsing into PBHs with $M$ at the formation time can be estimated as
\begin{equation}
\beta(M) = \int^{\infty}_{\delta_{\rm c}}{\rm d}\delta\; P(\delta) ~,
\label{eq:betam}
\end{equation}
where we use the critical value $\delta_{\rm c} = 0.4$ following Refs.~\cite{2009CQGra..26w5001M,2013PhRvD..88h4051H,2019PhRvL.122n1302G,2019arXiv190713311E}.
The mass of PBHs are related to the horizon mass at the formation time
as~(e.g. Ref.~\cite{2018PhRvD..97d3514I})
\begin{eqnarray}
M &=& \left. \gamma \rho \frac{4\pi H^{-3}}{3}\right|_{\eta = \eta_M}
 \notag \\
&\simeq&
\left( \frac{\gamma}{0.2} \right)
\left( \frac{g_{*}}{10.75} \right)^{-1/6}
\left( \frac{1}{\eta_M 1.9\times 10^{6} \;{\rm Mpc}^{-1}}\right)^{-2}
M_{\odot} ~,
\label{eq:pbhmass}
\end{eqnarray}
where 
$\eta_M$ is the formation time of the PBH with mass $M$, and $g_{*}$ and $\gamma$ are an effective degrees of freedom for energy density and a numerical factor which links the formed PBH mass with the horizon mass, respectively.

As already discussed in the previous section, the delta-function type power spectrum with the peak wavenumber $k_{\rm p}$ produce PBHs most effectively at $\eta_{\rm p}$.
The corresponding mass of those PBHs are obtained by substituting $\eta_{M} =\eta_{\rm p} = 0.987/k_{\rm p}$ into Eq.~\eqref{eq:pbhmass} as
\begin{equation}
M_{\rm p}(k_{\rm p}) = 
\left( \frac{\gamma}{0.2} \right)
\left( \frac{g_{*}}{10.75} \right)^{-1/6}
\left( \frac{k_{\rm p}}{1.88\times 10^{6} \;{\rm Mpc}^{-1}}\right)^{-2}
M_{\odot}
 ~.
\label{eq:mass-pk}
\end{equation}
In the followings, we will use this mass relation for each $k_{\rm p}$.
Denoting the density parameter of PBHs with mass $M$ over logarithmic
mass interval ${\rm d}\ln{M}$ by $\Omega_{\rm PBH}(M)$,
the current fraction of the total dark matter abundance in PBHs is given by~(e.g. Ref.~\cite{2018PhRvD..97d3514I})
\begin{eqnarray}
f(M)
&\equiv&
\frac{\Omega_{\rm PBH}}{\Omega_{\rm DM}}
\notag \\
&\simeq&
\left( \frac{\gamma}{0.2} \right)^{3/2}
\left( \frac{\beta(M)}{1.84\times 10^{-8}} \right)
\left( \frac{g_{*}}{10.75} \right)^{-1/4}
\left( \frac{\Omega_{\rm DM}h^{2}}{0.12} \right)^{-1}
\left( \frac{M}{M_{\odot}} \right)^{-1/2} ~.
\end{eqnarray}

\begin{figure}[t]
\centering
\includegraphics[width=0.6\textwidth]{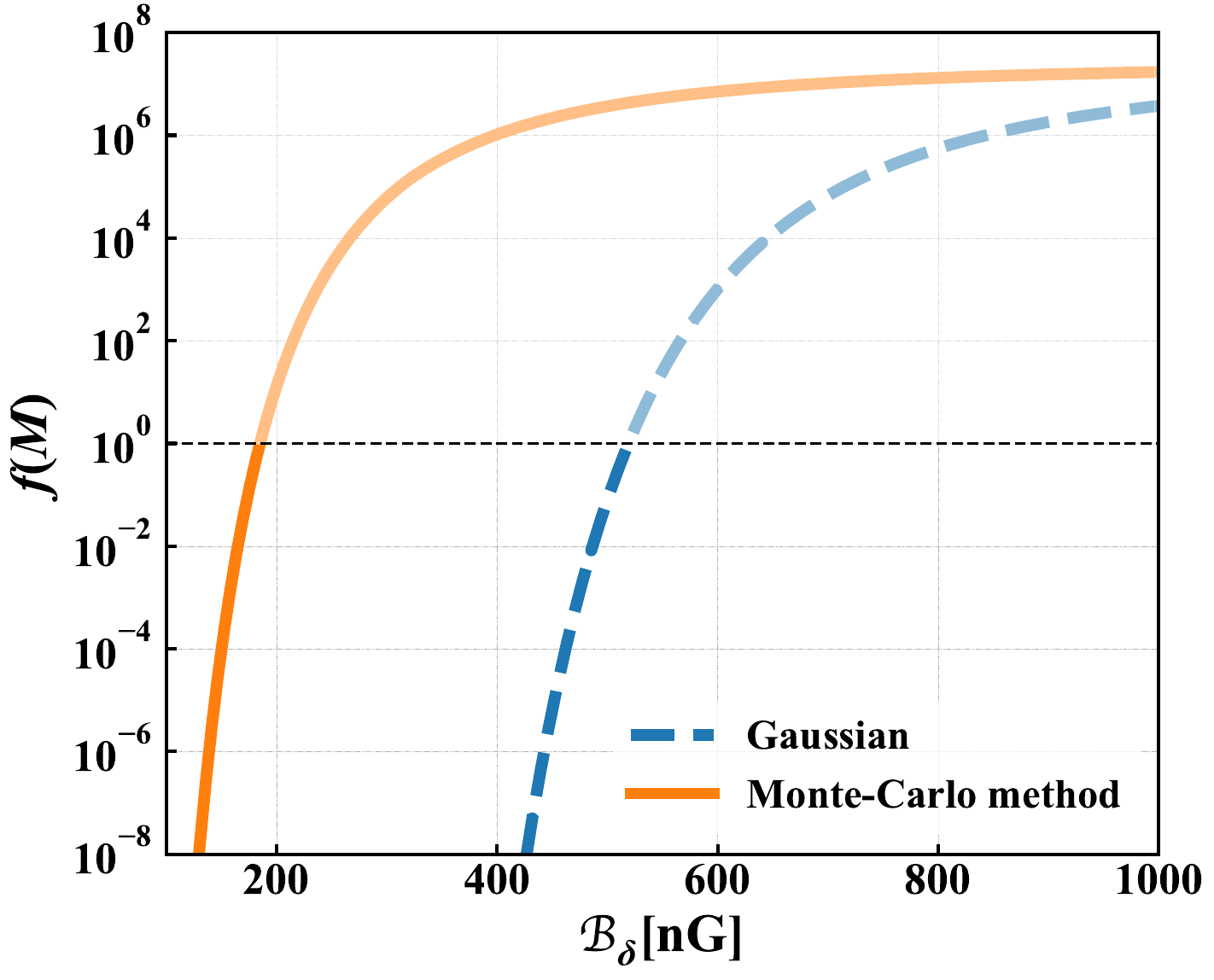}
\caption{
The amplitude of mass spectrum $f(M)$ for Gaussian case (blue dashed line) and non-Gaussian PDF obtained by the Monte-Carlo method case (orange solid line) with varying the amplitude of PMFs.
Black dashed line represents a unity.
}
\label{fig: mass spectrum}
\end{figure}

Now we obtain $f(M_{\rm p})$ for PBHs due to the passive mode of PMFs.
Substituting the PDF obtained from the Monte-Carlo method, $P_{\rm MC}$, given in the previous section \ref{sec: PDF} into Eq.~\eqref{eq:betam}, we obtain $f(M_{\rm p})$ including the non-Gaussian effect of the curvature perturbations.
In Fig.~\ref{fig: mass spectrum}, we show $f(M_{\rm p})$ as a function
of the amplitude of PMFs: ${\mathcal B}_\delta$.

To clarify the non-Gaussian effect induced by PMFs on the PBH formation, we show $f(M_{\rm p})$ obtained with a Gaussian PDF of the density perturbation as
\begin{equation}
P_{\rm G}(\delta) = \frac{1}{\sqrt{2\pi}\sigma}e^{-\frac{\delta^{2}}{2\sigma^{2}}} ~.\label{eq:gaussianP}
\end{equation}
In the case of the Gaussian PDF,
the fraction of the Universe collapsing into primordial black holes of
mass $M_{\rm p}$ is given with the complementary error function ${\rm erfc} (x)$ as
\begin{equation}
\beta_{\rm G}(M) = \int^{\infty}_{\delta_{\rm c}}{\rm d}\delta\; P_{\rm G}(\delta) = \frac{1}{2}{\rm erfc}\left( \frac{\delta_{\rm c}}{\sqrt{2}\sigma(M)}\right) ~,
\end{equation}
where we define the dispersion by using Eq.~(\ref{eq: dispersion}) as
\begin{eqnarray}
\sigma^{2}(M_{\rm p}) = \sigma^{2}_{\delta_{B}}(R=\eta_{\rm p},\eta_{\rm p}) ~.
 \label{eq:gaussiansig}
\end{eqnarray}
For comparison, we show the result with the Gaussian PDF as the dashed line
in Fig.~\ref{fig: mass spectrum}.
Since the non-Gaussian PDF obtained by the Monte-Carlo method has broad
distribution compared with the Gaussian PDF even for the same dispersion, the PBH abundance is
strongly enhanced than that with he Gaussian PDF.
In other words, when we use the Gaussian PDF given in Eq.~\eqref{eq:gaussianP} with the dispersion Eq.~\eqref{eq:gaussiansig}, the predicted abundance might be underestimated.
We find that, independently of $k_{\rm p}$, when the magnetic field amplitude
is larger than $\mathcal{B}_{\delta} \approx 184\; {\rm nG}$, the PBH fraction $f(M_{\rm p})$ exceeds unity.
Therefore, we obtain the simple but robust constraint on the delta-function type PMFs, $\mathcal{B}_{\delta} \lesssim 184\; {\rm nG}$ for all $k_{\rm p}$.

\subsection{Constraining PMFs from the PBH abundance}

Currently, $\beta(M)$ and/or $f(M)$ are well constrained by a number of observations.
We relate the upper limit on the PBH abundance with mass $M$ to that on the amplitude of PMFs with the peak wavenumber $k_{\rm p}$ which is given in terms of $M$ through Eq.~\eqref{eq:mass-pk}.
In this paper, we use the observed constraint on $\beta(M)$ and $f(M)$ as follows:
BBN constraint for $10^{-24}\lesssim M_{\rm PBH}/M_{\odot}\lesssim 10^{-18}$ \cite{Carr:2009jm},
the evaporating effect on CMB anisotropies for $10^{-20}\lesssim M_{\rm PBH}/M_{\odot}\lesssim 10^{-18}$ \cite{Carr:2009jm},
the evaporating effect of the light PBHs on the extragalactic gamma-ray for $10^{-19}\lesssim M_{\rm PBH}/M_{\odot}\lesssim 10^{-16}$ \cite{Carr:2009jm},
the femtolensing observation for $10^{-16}\lesssim M_{\rm PBH}/M_{\odot}\lesssim 10^{-14}$ \cite{2012PhRvD..86d3001B},
the microlensing observations with Subaru Hyper Supreme-Cam for $10^{-11}\lesssim M_{\rm PBH}/M_{\odot}\lesssim 10^{-5}$ \cite{2017arXiv170102151N,Niikura:2019kqi},
the microlensing observations with EROS/MACHO for $10^{-7}\lesssim M_{\rm PBH}/M_{\odot}\lesssim 10^{1}$ \cite{2007A&A...469..387T},
OGLE 5-year microlensing events for $10^{-6}\lesssim M_{\rm PBH}/M_{\odot}\lesssim 10^{-3}$ \cite{Niikura:2019kqi},
and the accretion effects on the CMB spectrum and temperature anisotropy for $10^{2}\lesssim M_{\rm PBH}/M_{\odot}\lesssim 10^{8}$ \cite{Carr:2009jm,2017PhRvD..95d3534A}.

\begin{figure}[t]
\centering
\includegraphics[width=0.8\textwidth]{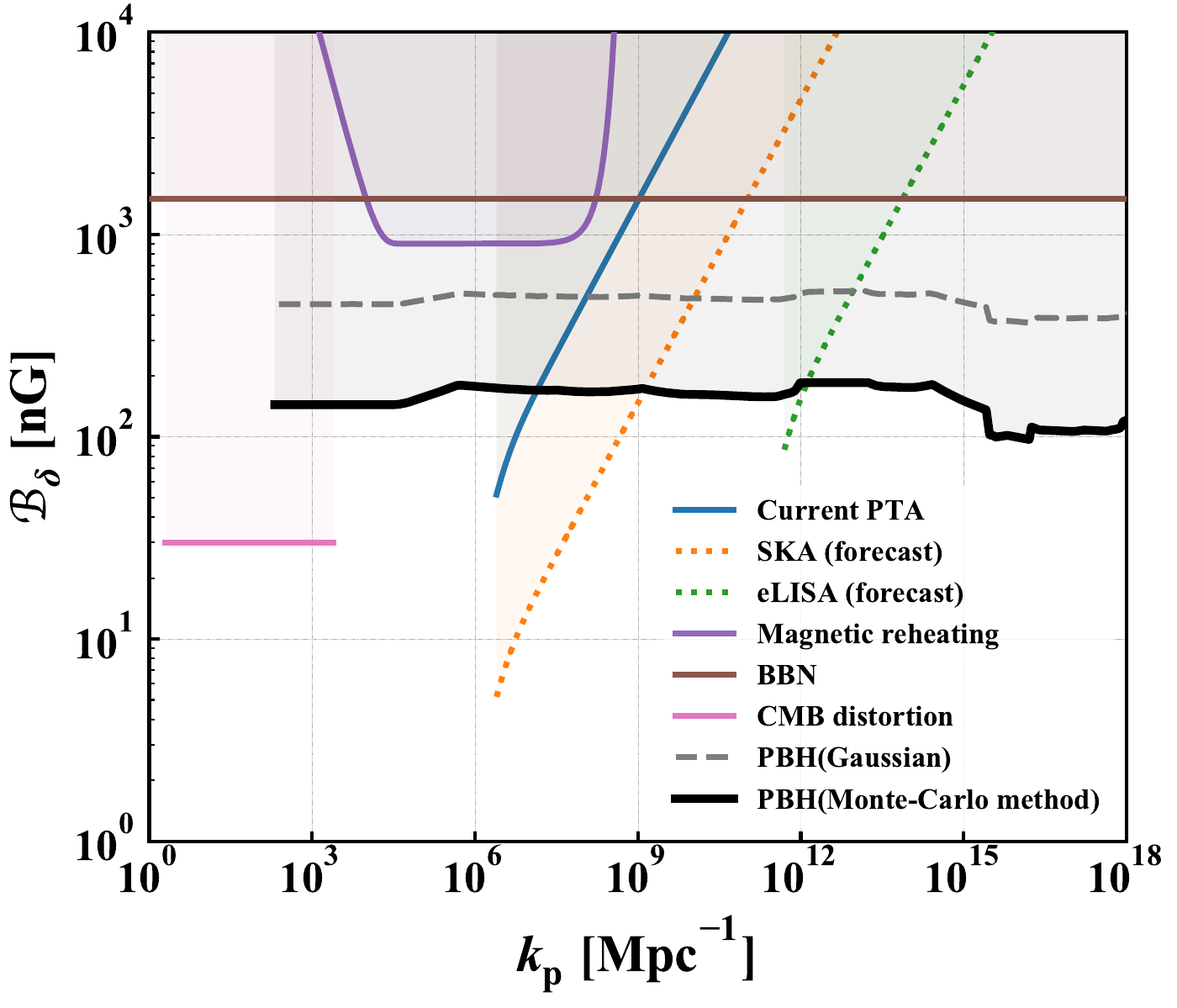}
\caption{
Upper limit on the amplitude of PMFs by using the available PBH constraints from Refs.~\cite{Carr:2009jm,2012PhRvD..86d3001B,2017arXiv170102151N,Niikura:2019kqi,2007A&A...469..387T,2017PhRvD..95d3534A} mentioned in the main text for mass range $10^{-24} \lesssim M/M_{\odot} \lesssim 10^{7}$.
Here, we fix $\eta_{B}/\eta_{\nu} = 10^{-12}$.
We also show the previous upper limits:
direct detection measurements of gravitational wave background by current and future PTAs such as SKA and eLISA \cite{2018PhRvD..98h3518S},
magnetic reheating \cite{Saga:2017wwr},
BBN \cite{2012PhRvD..86f3003K}
and CMB distortion \cite{2000PhRvL..85..700J}.
}
\label{fig: upper limit}
\end{figure}

We show the constraint on the amplitude of the delta-function type PMF power spectrum in Fig.~\ref{fig: upper limit}.
Note that on the PBH mass \textit{window} between $10^{-14} \lesssim M_{\rm PBH}/M_{\odot} \lesssim 10^{-11}$, we use the fact that the PBH abundance does not exceed the current total dark matter abundance, i.e., $f(M) \leq 1$.
For comparison, we present other current and future constraints on PMFs.
Whereas the direct detection measurements of gravitational wave background by PTAs (blue solid and orange dashed lines) provide the stronger constraints than the ones from PBHs, these mass ranges are very limited.
Currently the PBH constraint provides the tightest constraints on small-scale PMFs in the wide range of scales.

\section{Summary}\label{sec: summary}

In this paper, we investigated the primordial black hole (PBH) formation from the density perturbation generated by primordial magnetic fields (PMFs).
In the early universe, even on the super-horizon scales, the additional curvature perturbation is generated from the anisotropic stress of PMFs, which is called the passive mode.
The large amplitude of PMFs results in that of the curvature perturbation, or equivalently, that of the density perturbation.
If such density perturbation has sufficient amplitude to produce PBHs, we can exploit the upper limits on the PBH abundance to constrain the amplitude of PMFs.
In the standard picture of the PBH formation, we assume that the probability density function (PDF) follows the Gaussian statistics.
However, since the source of the density perturbation is the anisotropic stress of PMFs which is square of the magnetic field, in our case, the density perturbation must be non-Gaussian.
Therefore, we have to take into account the non-Gaussian effect for the PBH formation.

We carefully estimated the PDF by using the Monte-Carlo method based on Ref.~\cite{2016PhRvD..94d3507N}.
By assuming the delta-function type power spectrum for PMFs, we simulated the values of the density fluctuations in a thin spherical shell in the Fourier space.
Thereby, we found that the distribution of the density perturbation is a broader distribution than the Gaussian distribution.
According to the results of the Monte-Carlo simulation,
we can relate the amplitude of magnetic fields
to the current abundance of PBHs due to PMFs, 
including the non-Gaussian effect.

Finally, we found that, if the amplitude of PMFs is larger than $\mathcal{B}_{\delta} \approx 184 \; {\rm nG}$, the PBH abundance exceeds unity.
This upper limit is independent of the wave number of the PMF power spectrum.
We conclude that the amplitude of PMFs should not exceed this constrained value on all scales.
In addition, we used the observational upper limit on the PBH abundance which has been constrained on a broad mass range, i.e., $10^{-24}\lesssim M/M_{\odot}\lesssim 10^{7}$, and thereby we obtained the tighter constraint.
As a result, the amplitude of PMFs is constrained as a few hundred nano-Gauss on $10^{2}\;{\rm Mpc}^{-1} \leq k\leq 10^{18}\;{\rm Mpc}^{-1}$ shown in Fig.~\ref{fig: upper limit}.
These results are the strongest upper limit on small scales, where other cosmological observations must be insensitive.
The results might be helpful to distinguish the origin of magnetic fields from a number of magnetogenesis models.

\acknowledgments 
This work is supported by a Grant-in-Aid for Japan Society for Promotion of Science (JSPS) Research Fellow Number 17J10553~(S.S.), JSPS KAKENHI Grant Number 17H01110~(H.T.), and MEXT KAKENHI Grant Number 15H05888~(S.Y.) and 18H04356~(S.Y.).
Numerical computation in this work was carried out at the Yukawa Institute Computer Facility.

\appendix

\section{Probability density function of $\delta_{B}$}
\label{sec: app PDF}

Since the generated density perturbation from PMFs does not follow the Gaussian statistics, in order to discuss the PBH formation, we have to take into account the PDF of the generated density perturbation.
In this section, we obtain the PDF of the density perturbation by following Ref.~\cite{2016PhRvD..94d3507N}.

First of all, we decompose PMFs by using the polarization basis as
\begin{equation}
B_{i}(\bm{k}) = \sum_{r=+,\times}\epsilon^{(r)}_{i}(\hat{\bm{k}})B_{r}(\bm{k}) ~,
\end{equation}
where we define the polarization vectors with respect to the wave vector:
\begin{equation}
\hat{\bm{k}} =
\left(
\begin{array}{c}
\sin{\theta}\cos{\phi} \\
\sin{\theta}\sin{\phi} \\
\cos{\theta}
\end{array}
\right) ~,~~~
\bm{\epsilon}^{(+)}(\hat{\bm{k}}) =
\left(
\begin{array}{c}
\cos{\theta}\cos{\phi} \\
\cos{\theta}\sin{\phi} \\
-\sin{\theta}
\end{array}
\right)
~,~~~
\bm{\epsilon}^{(\times)}(\hat{\bm{k}}) =
\left(
\begin{array}{c}
- \sin{\phi} \\
\cos{\phi} \\
0
\end{array}
\right) ~.
\end{equation}
Then, the polarization vectors satisfy the followings
\begin{equation}
\bm{\epsilon}^{(+)}(-\hat{\bm{k}}) = \bm{\epsilon}^{(+)}(\hat{\bm{k}})
~,~~~
\bm{\epsilon}^{(\times)}(-\hat{\bm{k}}) = -\bm{\epsilon}^{(\times)}(\hat{\bm{k}})
~,~~~
\bm{\epsilon}^{(r)}(\hat{\bm{k}})\cdot\bm{\epsilon}^{(s)}(\hat{\bm{k}}) = \delta_{r,s} ~.
\end{equation}
In this notation, imposing the reality condition to PMFs, the Fourier modes of PMFs satisfy the following relations:
\begin{equation}
\left( B_{r}(\bm{k}) \right)^{*} = \eta_{r}B_{r}(-\bm{k}) ~,~~~
(\eta_{+} = +1 ~,~\eta_{\times} = -1)
~. \label{eq: reality condition b}
\end{equation}

Then the density perturbation defined in Eqs.~(\ref{eq:zetab}), (\ref{eq: def Pi_{B}}), and (\ref{eq: delta and zeta}) is given by 
\begin{equation}
\delta_{B}(\bm{k},\eta)
=
\sum_{r,s=+,\times}
\int\frac{{\rm d}^{3}k_{1}}{(2\pi)^{3}}
B_{r}(\bm{k}_{1})
B_{s}(\bm{k}-\bm{k}_{1})
G_{rs}(\eta, \bm{k},\bm{k}_{1})
~, \label{eq: Fourier delta}
\end{equation}
where we define
\begin{equation}
G_{rs}(\eta, \bm{k},\bm{k}_{1})
= 
\frac{\xi R_{\gamma}}{6\pi \rho_{\gamma,0}}
\left( \frac{k}{aH}\right)^{2}
\left( \hat{k}_{i}\hat{k}_{j} - \frac{1}{3}\delta_{ij}\right)
\epsilon^{(r)}_{i}(\hat{\bm{k}}_{1})\epsilon^{(s)}_{j}(\widehat{\bm{k}-\bm{k}_{1}})
~.
\end{equation}
Note that $G_{rs}(\eta, \bm{k},\bm{k}_{1})$ satisfies
\begin{equation}
G_{rs}(\eta, -\bm{k},-\bm{k}_{1}) =
\eta_{r}\eta_{s}G_{rs}(\eta, \bm{k},\bm{k}_{1}) ~.
\end{equation}

The expression of Eq.~(\ref{eq: Fourier delta}) in real space with smoothing over the radius $R$ is given by
\begin{equation}
\delta_{B}(\eta, \bm{x}=\bm{0}; R)
=
\sum_{r,s=+,\times}
\int\frac{{\rm d}^{3}k_{1}}{(2\pi)^{3}}
\int\frac{{\rm d}^{3}k_{2}}{(2\pi)^{3}}
W(|\bm{k}_{1}+\bm{k}_{2}|R)
B_{r}(\bm{k}_{1})
B_{s}(\bm{k}_{2})
G_{rs}(\eta, \bm{k}_{1}+\bm{k}_{2},\bm{k}_{1}) ~.
\end{equation}
The corresponding discrete expression is given by
\begin{equation}
\delta_{B}(\eta, \bm{x}=\bm{0}; R)
= \frac{({\rm d}k)^{6}}{(2\pi)^{6}}
\Biggl[
\sum_{r,s=+,\times}\sum_{\bm{k}_{i},\bm{k}_{j}\in S}
W(|\bm{k}_{i}+\bm{k}_{j}| R)
B_{r}(\bm{k}_{i})
B_{s}(\bm{k}_{j})
G_{rs}(\eta, \bm{k}_{i}+\bm{k}_{j}, \bm{k_{i}})
\Biggr] ~, \label{eq: delta discrete}
\end{equation}
where $S$ is the set of the grid points inside the spherical shell because we are focusing on the delta-function type power spectrum given in Eq.~(\ref{eq: PS def}).

Here, we decompose the Fourier component $B_{r}(\bm{k})$ as following:
\begin{equation}
B_{r}(\bm{k})
= a_{r}(\bm{k}) + i b_{r}(\bm{k}) ~,
\end{equation}
where $a_{r}(\bm{k})$ and $b_{r}(\bm{k})$ are real Gaussian random variables, which satisfy the following conditions by using Eq.~(\ref{eq: reality condition b}) as
\begin{equation}
a_{+}(\bm{k}) = a_{+}(-\bm{k}) ~,~~~
b_{+}(\bm{k}) = - b_{+}(-\bm{k}) ~,~~~
a_{\times}(\bm{k}) = -a_{\times}(-\bm{k}) ~,~~~
b_{\times}(\bm{k}) = b_{\times}(-\bm{k}) ~.
\end{equation}
The dispersion of $a_{r}$ and $b_{r}$ is given by
\begin{equation}
\sigma^{2} = \frac{\pi^{2}}{k^{3}_{\rm p}}\frac{1}{\epsilon}\left( \frac{{\rm d}k}{2\pi} \right)^{-3} ~,
\end{equation}
where ${\rm d}k = \epsilon k_{\rm p}$ is the interval between two neighboring grid points and $\epsilon$ determines the thickness of the spherical shell.
We use $\epsilon = 0.05$, and we check that this value gives converged result.

Then, by performing same calculation in Ref.~\cite{2016PhRvD..94d3507N}, we obtain the density fluctuation in the discrete expression as
\begin{eqnarray}
\delta_{B}(\eta, \bm{x}=\bm{0}; R)
&=&
\frac{({\rm d}k)^{3}}{8\pi \epsilon k^{3}_{\rm p}}
\Biggl[
\bm{a}^{\rm t}\bm{M}^{a}\bm{a} + \bm{b}\bm{M}^{b} \bm{b}
\Biggr]~ , \label{eq: delta to be simulated}
\end{eqnarray}
where $\bm{a}$ and $\bm{b}$ are the $2N$ dimensional vectors, and $\bm{M}^{a/b}$ is the $2N\times 2N$ matrix, which are defined as follows:
\begin{eqnarray}
\bm{M}^{a/b} &=&
\left(
\begin{array}{cc}
\bm{M}^{a/b}_{++} & \bm{M}^{a/b}_{+\times} \\
\bm{M}^{a/b}_{\times+} & \bm{M}^{a/b}_{\times\times} \\
\end{array}
\right) \label{eq: matrix}
~,\\
\bm{a}^{\rm t}
&=& \sigma^{-1}
\left(
a_{+}(\bm{k}_{1}), a_{+}(\bm{k}_{2}), \cdots , a_{+}(\bm{k}_{N}), 
a_{\times}(\bm{k}_{1}), a_{\times}(\bm{k}_{2}) , \cdots , a_{\times}(\bm{k}_{N})
\right)
~,\\
\bm{b}^{\rm t}
&=& \sigma^{-1}
\left(
b_{+}(\bm{k}_{1}), b_{+}(\bm{k}_{2}), \cdots , b_{+}(\bm{k}_{N}), 
b_{\times}(\bm{k}_{1}), b_{\times}(\bm{k}_{2}) , \cdots , b_{\times}(\bm{k}_{N})
\right) ~,
\end{eqnarray}
and
\begin{align}
M^{a}_{rs}(\bm{k}_{i},\bm{k}_{j}) &=
2\Bigl(
W(|\bm{k}_{i}+\bm{k}_{j}| R)G_{rs}(\eta, \bm{k}_{i}+\bm{k}_{j}, \bm{k_{i}})
+ \eta_{s} W(|\bm{k}_{i}-\bm{k}_{j}| R) G_{rs}(\eta, \bm{k}_{i}-\bm{k}_{j}, \bm{k_{i}})
\Bigr) ~, \\
M^{b}_{rs}(\bm{k}_{i},\bm{k}_{j}) &=
2 \Bigl(
- W(|\bm{k}_{i}+\bm{k}_{j}| R)G_{rs}(\eta, \bm{k}_{i}+\bm{k}_{j}, \bm{k_{i}})
+ \eta_{s} W(|\bm{k}_{i}-\bm{k}_{j}| R) G_{rs}(\eta, \bm{k}_{i}-\bm{k}_{j}, \bm{k_{i}})
\Bigr) ~.
\end{align}
Note that $N$ is the total number of grid points inside the spherical shell and in the current parameter settings, $4N\approx 10^{4}$.

Finally, we diagonalize the $4N\times 4N$ matrix and the density perturbation is formally written as
\begin{equation}
\delta_{B} = \sum_{i=1}^{4N}a_{i} x^{2}_{i} ~, \label{eq: delta B discrete}
\end{equation}
where $a_{i}$ are the eigenvalues of the matrix and $x_{i}$ are the independent Gaussian random variables with dispersion being unity.
In order to obtain the PDF of the density perturbation, we use the Monte-Carlo method to Eq.~(\ref{eq: delta B discrete}) by diagonalizing the $4N\times 4N$ matrix Eq.~(\ref{eq: matrix}).

\bibliographystyle{JHEP}
\bibliography{ref}
\end{document}